\documentclass[12pt]{article}
\usepackage{amsmath}
\begin{document}
\begin{center}
\LARGE
\textbf{Why Should we Interpret\\
Quantum Mechanics?}\\[1cm]
\large
\textbf{Louis Marchildon}\\[0.5cm]
\normalsize
D\'{e}partement de physique,
Universit\'{e} du Qu\'{e}bec,\\
Trois-Rivi\`{e}res, Qc.\ Canada G9A 5H7\\
email: marchild$\hspace{0.3em}a\hspace{-0.8em}
\bigcirc$uqtr.ca\\
\end{center}
\medskip
\begin{abstract}
The development of quantum information
theory has renewed interest in the
idea that the state vector
does not represent the state of a quantum
system, but rather the knowledge
or information that we may have on the
system.  I argue that this epistemic
view of states appears to solve foundational
problems of quantum mechanics only at
the price of being essentially incomplete.
\end{abstract}
\section{INTRODUCTION}
The foundations and the interpretation of
quantum mechanics, much discussed by the
founders of the theory, have blossomed anew
in the past few decades.  It has been
pointed out that every single year between 1972 and
2001 has witnessed at least one scientific meeting
devoted to these issues~\cite{fuchs}.

Two problems in the foundations of quantum
mechanics stand out prominently.  The first one
concerns the relationship between quantum mechanics
and general relativity.  Both theories are highly
successful empirically, but they appear to be
mutually inconsistent.  Most investigators
used to believe that the problem lies more with
general relativity than with quantum mechanics, and
that the solution would consist in coming up with
a suitable quantum version of Einstein's
gravitational theory.  More recently, however,
it has become likely that a correct quantum
theory of gravity may involve leaps
substantially bolder~\cite{smolin}.

The second problem is more down-to-earth, and
shows up starkly in ordinary nonrelativistic
quantum mechanics.  It consists in the reconciliation
of the apparently indeterminate nature of quantum
observables with the apparently determinate
nature of classical observables, and it crystallizes
in the so-called measurement problem.
Broadly speaking, there
are two ways to address it.  The first one, elaborated
by von Neumann~\cite{neumann}, consists in denying
the universal validity of the unitary evolution of
quantum states, governed by the Schr\"{o}dinger
equation.  State vector collapse, postulated by
von Neumann but not worked out in detail by him,
was made much more precise recently in spontaneous
localization models~\cite{ghirardi}.  The second
road to reconciliation consists in stressing the
universal validity of the Schr\"{o}dinger
equation, while assigning values to specific
observables of which the state vector is not
necessarily an eigenstate.  Bohmian
mechanics~\cite{bohm1,bohm2,holland} and
modal interpretations~\cite{fraassen1,vermaas},
among others, fall in this category.

The past twenty years have also witnessed the
inception and quick development of quantum
information theory.  This was not unrelated to
foundational studies.  Quantum
cryptography~\cite{bennett} and quantum
teleportation~\cite{brassard}, for instance, are
based on the Einstein-Podolsky-Rosen setup
and algorithms for fast computation use
entanglement in an essential way~\cite{nielsen}.

As foundational studies contributed to
quantum information theory, a number of
investigators feel that quantum information
theory has much to contribute to the interpretation
of quantum mechanics.  This has to do with what
has been called the \emph{epistemic view} of
state vectors, which goes back at least to
Heisenberg~\cite{heisenberg} but has
significantly evolved in the past few
years~\cite{fuchs,peierls,rovelli,peres,spekkens,bub}.
In the epistemic view, the state vector (or wave
function, or density matrix) does not represent
the objective state of a microscopic system
(like an atom, an electron, a photon), but
rather our knowledge of the probabilities
of outcomes of macroscopic measurements.
This, so the argument goes, considerably
clarifies, or even completely dissolves,
the EPR paradox and the measurement problem.

The purpose of this paper is to examine the
status of the epistemic view.  It is a minimal
interpretation, and indeed was referred to as an
``interpretation without interpretation''~\cite{peres}.
The question of the adequacy of the epistemic
view is much the same as the one whether
quantum mechanics needs being interpreted.

In Sec.~2, I shall review the main arguments
in favor of the epistemic view of quantum states.
They have to do with state vector collapse and the
consistency of quantum mechanics with special
relativity.  It turns out that the issue
of interpretation can fruitfully be
analyzed in the context of the \emph{semantic view}
of theories.  As outlined in Sec.~3, this
approach tries to answer the question, How
can the world be the way the theory says it is?  The
next section will describe a world where the epistemic
view would be adequate.  This, it turns out, is not
the world we live in, and Secs.~5 and~6 will argue that
in the real world, the epistemic view in fact
begs the question of interpretation.
%
%\newpage
\begin{sloppypar}
\section{ARGUMENTS FOR THE EPISTEMIC VIEW}
\end{sloppypar}
Interpretations of quantum mechanics can be
asserted with various degrees of
illocutionary strength.  The many-worlds
interpretation~\cite{everett}, for instance,
has sometimes been presented as following directly
from the quantum-mechanical formalism.
In a similar vein, Rovelli has proposed that
the epistemic view follows from the observation
that different observers may give different
accounts of the same
sequence of events~\cite{rovelli}.

Rovelli's argument goes as follows.  A quantum
system $S$ with two-dimensional state space is
initially described by the state vector
$|\psi\rangle = \alpha |1\rangle + \beta |2\rangle$,
where $|1\rangle$ and $|2\rangle$ are orthonormal
eigenvectors of an observable $Q$.  Sometime
between $t_i$ and $t_f$, an apparatus $O$ (which
may or may not include a human being)
performs a measurement of $Q$ on $S$, obtaining
the value 1.  Call this experiment $\mathcal{E}$.  
According to standard quantum mechanics, $O$
describes $\mathcal{E}$ as
\begin{align}
t_i & \longrightarrow t_f \notag \\
\alpha |1\rangle + \beta |2\rangle
& \longrightarrow |1\rangle . \label{rov1}
\end{align}

Now let a second observer $P$ describe the
compound system made up of $S$ and $O$.
Let $|O_i\rangle$ be the state vector of $O$
at $t_i$.  The measurement interaction between
$O$ and $S$ implies that their total state
vector becomes entangled.  Thus from the point
of view of $P$, who performs no measurement
between $t_i$ and $t_f$, the experiment
$\mathcal{E}$ should be described as
\begin{align}
t_i & \longrightarrow t_f \notag \\
(\alpha |1\rangle + \beta |2\rangle) \otimes |O_i\rangle
& \longrightarrow \alpha |1\rangle \otimes |O1\rangle
+ \beta |2\rangle \otimes |O2\rangle . \label{rov2}
\end{align}
Here $|O1\rangle$ and $|O2\rangle$ are pointer
states of $O$ showing results 1 and 2.

According to Rovelli, Eq.~(\ref{rov1}) is the
conventional quantum-mechanical description
of experiment $\mathcal{E}$ from the point of
view of $O$, whereas Eq.~(\ref{rov2})
is the description of
$\mathcal{E}$ from the point of view of $P$.
From this he concludes that ``[i]n quantum mechanics
different observers may give different accounts of
the same sequence of events.''  Or, to borrow the
title of his Sec.~2, ``Quantum mechanics is a
theory about information.''

To someone who believes there is a state of
affairs of some sort behind the description, the
difference between O's and P's point of view
means that one of them is mistaken.  To put
it differently, the
problem with the argument is that expressions
like ``standard quantum mechanics'' or
``conventional quantum-mechanical description''
are ambiguous.  They can refer
either (i) to strict unitary Schr\"{o}dinger evolution,
or (ii) in the manner of von Neumann, to
Schr\"{o}dinger evolution \emph{and} collapse.
Once a precise definition is agreed upon, either 
description~(\ref{rov1}) or description~(\ref{rov2})
(but not both) is correct.  Strict
Schr\"{o}dinger evolution (i) is encapsulated
in Eq.~(\ref{rov2}), whereas Schr\"{o}dinger
evolution with collapse (ii) is encapsulated
in Eq.~(\ref{rov1}).  As one may see it,
the issue is not that
the description depends on the observer,
but that with any precise definition there
is a specific problem.  With
definition (i), quantum mechanics seems to
contradict experiment, while with definition (ii),
one needs to provide an explicit collapse
mechanism.  Thus Rovelli's discussion, rather than
establishing the epistemic view, brings us back
to the foundational problem outlined in
Sec.~1.\footnote{In private exchanges, Rovelli
stressed that the relational character of quantum
theory is the solution of the apparent
contradiction between discarding Schr\"{o}dinger
evolution without collapse and not providing
an explicit collapse mechanism.
My criticism of its starting
point notwithstanding, much of the discussion
in Ref.~\cite{rovelli} is lucid and
thought-provoking.}

The epistemic view can also be propounded not
as a consequence of the quantum-mechanical
formalism, but as a way to solve
the foundational problems.  It does so by denying
that the (in this context utterly misnamed) state
vector represents the state of a microscopic
system.  Rather, the state vector represents
knowledge about the probabilities of results of
measurements performed in a given context with a
macroscopic apparatus, that is, information about
``the potential consequences of our experimental
interventions into nature''~\cite{fuchs}.

How does the epistemic view deal with the
measurement problem?  It does so by construing
the collapse of the state vector not as a
physical process, but as a change of
knowledge~\cite{peierls}.  Insofar as the state
vector is interpreted as objectively describing
the state of a physical system, its abrupt change
in a measurement
implies a similar change in the system, which
calls for explanation.  If, on the other hand,
the state vector describes knowledge of
conditional probabilities (i.e.\ probabilities of
future macroscopic events conditional on past
macroscopic events), then as long as what is 
conditionalized upon remains the same, the state
vector evolves unitarily.  It collapses when the
knowledge base changes, thereby simply reflecting
the change in the conditions being held fixed
in the specification of probabilities.

The epistemic view also helps in removing the
clash between collapse and Lorentz
invariance~\cite{fuchs,bloch}.  Take for
instance the EPR setup with two spin 1/2
particles, and let the state vector $|\chi\rangle$
of the compound system be an eigenstate of the
total spin operator with eigenvalue zero.  One
can write
\begin{equation}
|\chi\rangle = \frac{1}{\sqrt{2}}
\left\{ |+; \mathbf{n} \rangle \otimes |-; \mathbf{n} \rangle -
|-; \mathbf{n} \rangle \otimes |+; \mathbf{n} \rangle \right\} ,
\end{equation}
where the first vector in a tensor product refers to
particle $A$ and the second vector to particle $B$.
The vector $|+; \mathbf{n} \rangle $, for instance,
stands for an eigenvector of the $\mathbf{n}$-component
of the particle's spin operator, with eigenvalue
$+1$ (in units of $\hbar/2$).
The unit vector $\mathbf{n}$ can point in any direction,
a freedom which corresponds to the
rotational symmetry of $|\chi\rangle$.

Suppose Alice measures the $\mathbf{n}$-component
of $A$'s spin and obtains the value $+1$.  If
the state vector represents the objective state of
a quantum system and if collapse is a physical
mechanism, then $B$'s state immediately collapses
to $|-; \mathbf{n}\rangle$.  This explains why
Bob's subsequent measurement of the
$\mathbf{n}$-component of $B$'s spin yields the
value $-1$ with certainty.  Thus Alice's choice
of axis at once determines the possible states
into which $B$ may collapse, and Alice's result
immediately singles out one of these states.

If the two measurements are spacelike separated,
there exist Lorentz frames where Bob's
measurement is earlier
than Alice's.  In these frames the instantaneous
collapse is triggered by Bob's measurement.
This ambiguity, together with the fact that what
is instantaneous in one frame is not in others,
underscores the difficulty of reconciling
relativistic covariance with physical collapse.

In the epistemic view, what changes when Alice
performs a measurement is Alice's knowledge.
Bob's knowledge will change either if he himself
performs a measurement, or if Alice sends him
the result of her measurement by conventional
means.  Hence there is no physical collapse
on a spacelike hypersurface.

To the proponents of the epistemic view,
the above arguments show that it considerably
attenuates, or even completely solves, the
problems associated with quantum measurements
and collapse.  I will argue, however, that this
result is achieved only at the price of giving up
the search for a spelled out consistent view
of nature.\footnote{The epistemic view was
criticized from a different perspective by
Zeh~\cite{zeh}.}
\bigskip
%
%\newpage
\section{THE SEMANTIC VIEW OF THEORIES}
Investigations on the structure of scientific
theories and the way they relate to phenomena
have kept philosophers of science busy for
much of the twentieth century.  In the past
few decades, the semantic view has emerged
as one of the leading approaches to these
problems~\cite{fraassen1,giere,suppe,fraassen2}.
Among other issues it helps clarifying,
I believe it sheds considerable light
on the question of the
interpretation of quantum mechanics.

In the semantic view a scientific theory
is a structure, defined primarily by models
rather than by axioms or a specific linguistic
formulation.  A general theoretical
hypothesis then asserts
that a class of real systems, under suitable
conditions of abstraction and idealization,
belongs to the class of models entertained
by the theory.  If the theory is empirically
adequate, the real systems behave (e.g.\ evolve
in time) in a way predictable on the basis
of the models.  Yet the empirical agreement
may leave considerable room on the way the
world can be for the theory to be true.  This
is the question of interpretation.

This succinct characterization can best
be understood by means of examples.  Take
classical particle mechanics.  The (mathematical)
structure consists of constants $m_i$, functions
$\mathbf{r}_i (t)$, and vector fields $\mathbf{F}_i$
(understood as masses, positions, and forces),
together with the system of second-order differential
equations $\mathbf{F}_i = m_i \mathbf{a}_i$.  A
particular model is a system of ten point masses
interacting through the $1/r^2$ gravitational
force.  A theoretical hypothesis then
asserts that the solar system corresponds to
this model, if the sun and nine planets are
considered pointlike and all other objects
neglected.  Predictions made on the basis of
the model correspond rather well with reality,
especially if suitable correction factors are
introduced in the process of abstraction.
But obviously the model can be made much more
sophisticated, taking into account for instance
the shape of the sun and planets, the planets'
satellites, interplanetary matter, and so on. 

Now what does the theory have to say about
how a world of interacting masses is really like?
It turns out that such a world can be viewed
in (at least) two empirically equivalent but
conceptually very different ways.  One can assert
that it is made only of small (or extended) masses
that interact by instantaneous action at a distance.
Or else, one can say that the masses produce
everywhere in space a gravitational field, which
then locally exerts forces on the masses.  These
are two different interpretations of the theory.
Each one expresses a possible way of making the
theory true (assuming empirical adequacy).
Whether the world is such that masses instantaneously
interact at a distance in a vacuum, or a
genuine gravitational field is produced throughout
space, the theory can be held as truly realized.

A similar discussion can be made with
classical electromagnetism.  The mathematical
structure consists of source
fields $\rho$ and $\mathbf{j}$
(understood as charge and current densities),
and vector fields $\mathbf{E}$ and $\mathbf{B}$
(electric and magnetic fields) related to the
former through Maxwell's equations.  A function
$\mathbf{F}$ of the fields (the Lorentz force)
governs the motion of charge and current
carriers.  A model of the theory might
be a perfectly conducting rectangular guide
with a cylindrical dielectric post, subject
to an incoming TE$_{10}$ wave.

Again, each specific model of interacting
charges and currents can be viewed in empirically
equivalent and conceptually different ways.
One can allow only retarded fields, or both
retarded and advanced fields~\cite{wheeler}.
If one assumes the existence of a complete
absorber, one can get rid of the fields
entirely, and view electrodynamics as a theory
of moving charges acting
on each other at a distance (though not
instantaneously).  Although the interpretation
with genuine retarded fields is the one by far
the more widely accepted, the other interpretations
also provide a way by which Maxwell's theory can be
true.

Let us now turn to quantum mechanics.  The
mathematical structure consists of state spaces
$\mathcal{H}$, state vectors $|\psi\rangle$, and
density operators $\rho$; of Hermitian operators
$A$, eigenvalues $|a\rangle$, and eigenvectors $a$;
of projectors $P_a$, etc.  Defining quantum mechanics
as covering strict unitary evolution, state
vectors evolve according to the Schr\"{o}dinger
equation.  The scope of the theory is specified
(minimally) by associating eigenvalues $a$ to results
of possible measurements, and quantities like
$\mbox{Tr} (\rho P_a)$ to corresponding probabilities.

One kind of real system that can be modelled
on the basis of the structure is
interference in a two-slit Young setup.  Often
the system can conveniently be restricted to the
$xy$ plane.  In a specific model
the source can be represented by
a plane wave moving in the $x$ direction, and the
slits can be taken as modulating gaussian wave
packets in the $y$ direction.  These packets then
propagate according to the Schr\"{o}dinger
equation in free space and produce a wave
pattern on a screen.  The absolute square of
the wave amplitude is associated with the
probability that a suitable detector will
click, as a function of its position on
the screen.

From the semantic point of view, the question
of interpretation is the following: How can the
world be so that quantum models representing
Young setups (as well as other situations)
are empirically adequate?  The Copenhagen answer
(or at least a variant of it) says that
micro-objects responsible for the fringes don't
have well-defined properties unless these are
measured, but that large scale apparatus always have
well-defined properties.  I share the view
of those who believe that this answer is complete only
insofar as it precisely specifies the transition
scale between the quantum and the classical.  Bohm's
answer to the above question is that all
particles always have precise positions,
and these positions are the ones that
show up on screens~\cite{philipidis}.  The
many-worlds view is that different
detector outcomes simultaneously exist in
different worlds or in different minds.  I shall
shortly attempt to assess the adequacy of
the epistemic view, but first we should look
at a world especially tailored to such an
interpretation.
%
%\newpage
\section{A WORLD FOR THE EPISTEMIC VIEW}
Consider a hypothetical world where large
scale objects (meaning objects much larger
than atomic sizes) behave, for all practical
purposes, like large scale objects in the
real world.  The trajectories of javelins and
footballs can be computed accurately by means
of classical mechanics with the use of a
uniform downward force and air resistance.
Waveguides and voltmeters obey
Maxwell's equations.  Steam engines and air
conditioners work according to the laws of
classical thermodynamics.
The motion of planets is well described by
Newton's laws of gravitation and of motion,
slightly corrected by the equations of
general relativity.

Close to atomic scales, however, these laws
may no longer hold.  Except for one restriction
soon to be spelled out, I shall not be specific
about the changes that macroscopic laws may or
may not undergo in the microscopic realm.  Matter, for
instance, could either be continuous down to
the smallest scales, or made of a small
number of constituent particles like our
atoms.  The laws of particles and fields
could be the same at all scales, or else
they could undergo significant changes
as we probed smaller and smaller distances.

In the hypothetical world one can perform
experiments with pieces of equipment like 
Young's two-slit setup and Stern-Gerlach
devices.  The Young type experiment, for instance,
uses two macroscopic devices $E$ and $D$ that
both have on and off states and work in the
following way.  Whenever $D$ is suitably
oriented with respect to $E$ (say, roughly
along the $x$ axis) and both are in
the on state, $D$ clicks in a more or less
random way.  The average time interval
between clicks depends on the distance $r$
between $D$ and $E$, and falls roughly as
$1/r^2$.  The clicking stops if a shield of a
suitable material is placed perpendicularly
to the $x$ axis, between $D$ and $E$.

If holes are pierced through the shield, however,
the clicking resumes.  In particular, with
two small holes of appropriate size and
separation, differences in the clicking rate
are observed for small transverse displacements
of $D$ behind the shield.  A
plot of the clicking rate against $D$'s
transverse coordinate displays maxima and
minima just as in a wave interference pattern.
No such maxima and minima are observed, however,
if just one hole is open or if both holes
are open alternately.

At this stage everything happens as if $E$
emitted some kind of particles and $D$
detected them, and the particles behaved
according to the rules of quantum mechanics.
Nevertheless, we shall nor commit ourselves
to the existence or nonexistence of these
particles, except on one count.  Such particles,
if they exist, are not in any way related to
hypothetical constituents of the material
making up $D$, $E$, or the shield, or of any
macroscopic object whatsoever.  Whatever
the microscopic structure of macroscopic
objects, it has nothing to do with what is
responsible for the correlations between
$D$ and $E$.\footnote{Any objection to
the hypothetical world on grounds that
energy might not be conserved, action would
not entail corresponding reaction, and so on,
would miss the point, which is to clarify in what
context the epistemic view can be held
appropriately.}

In the hypothetical world one can also perform
experiments with Stern-Gerlach setups.
Again a macroscopic device $D'$ clicks more
or less randomly when suitably oriented
with respect to another macroscopic device
$E'$, both being in the on state.
If a large magnet (producing a strongly
inhomogeneous magnetic field) is placed
along the $x$ axis between $E'$ and 
$D'$, clicks are no longer observed
where $D'$ used to be.  Rather they are observed
if $D'$ is moved transversally to (say) two
different positions, symmetrically oriented
with respect to the $x$ axis.  Let $\xi$
be the axis going from the first magnet
to one of these positions.  If a second
magnet is put behind the first one
along the $\xi$ axis and $D'$ is placed
further behind, then $D'$ clicks in two
different positions in the plane
perpendicular to $\xi$.  In the
hypothetical world the
average clicking rate depends on the magnet's
orientation, and it follows the
quantum-mechanical rules of spin 1/2 particles.
Once again, however, we assume that the
phenomenon responsible for the correlations
between $D'$ and $E'$ has nothing to do with
hypothetical constituents of macroscopic
objects.

In the two experiments just described,
quantum mechanics correctly predicts
the correlations between $D$ and $E$
(or $D'$ and $E'$) when suitable
macroscopic devices are interposed
between them.  In that situation, the
theory can be interpreted in (at least) two
broadly different ways.  In one interpretation,
the theory is understood as applying to
genuine microscopic objects, emitted by $E$
and detected by $D$.  Perhaps these objects
follow Bohmian like trajectories, or behave
between emitter and detector in some other
way compatible with quantum mechanics.  In the
other interpretation, there are no microscopic
objects whatsoever going from $E$ to $D$.
There may be something like an action at a distance.
At any rate the theory is in that case
interpreted instrumentally,
for the purpose of quantitatively
accounting for correlations in the stochastic
behavior of $E$ and $D$.

In the hypothetical world being
considered, I would maintain
that both interpretations
are logically consistent and adequate.  Both clearly
spell out how the world can possibly be the way
the theory says it is.  Of course, each
particular investigator
can find more satisfaction in one
interpretation than in the other.  The
epistemic view corresponds here to the
instrumentalist interpretation.  It
simply rejects the existence of
microscopic objects that have no other use
than the one of predicting observed
correlations between macroscopic objects.
%
%\newpage
\section{INSUFFICIENCY OF THE EPISTEMIC VIEW}
We shall assume that macroscopic objects
exist and are always in definite
states.\footnote{By this assumption, I mean
that they are not in quantum superpositions
of macroscopically distinct states.  They may
still be subject to the very tiny uncertainties
required by Heisenberg's principle.}  Not everyone
agrees with this.  Idealistic thinkers believe
there is nothing outside mind, and some
fruitful interpretations of quantum mechanics
(like the many-minds interpretation) claim we
are mistaken in assuming the definiteness of
macroscopic states.

For the purpose of asserting the relevance
of the epistemic view, however, these assumptions
can be maintained.  Indeed much of the appeal
of the epistemic view is that it appears
to reconcile them with the exact validity of
quantum mechanics.

All scientists today believe that macroscopic
objects are in some sense made of atoms and
molecules or, more fundamentally, of electrons,
protons, neutrons, photons, etc.  The epistemic
view claims that state vectors do not represent
states of microscopic objects, but knowledge of
probabilities of experimental results.  I suggest
that with respect to atoms, electrons, and
similar entities this can mean broadly
either of three things:
\begin{enumerate}
\item Micro-objects do not exist~\cite{ulfbeck}.
\item Micro-objects may exist but they have no states.
\item Micro-objects may exist and may have states, but
attempts at narrowing down their existence or
specifying their states are useless, confusing, or
methodologically inappropriate.
\end{enumerate}

In the first case, the question that immediately
comes to mind is the following: How can
something that exists (macroscopic objects)
be made of something that does not exist
(micro-objects)?\footnote{I can find no answer
to this question in Ref.~\cite{ulfbeck} which,
however, fits remarkably well with the
hypothetical world of Sec.~4.}
And in the second case, we
can ask similarly: How can something that has a
state be made of something that does not have one?

Can we conclude from these interrogations
that the epistemic view is logically inconsistent?  
Is the argument of the last paragraph
a \emph{reductio ad absurdum}?  Not so.  What the
questions really ask is the following:
How is it possible that the world be like that?
How, for instance, can we have a well-defined
macroscopic state starting from objects that do
not have states?  This, we have seen, is
precisely the question of interpretation.  Hence
if the epistemic view is asserted as in cases
(1) and (2), our discussion shows that it is
incomplete and paradoxical, and that the process
of completion and paradox resolution coincides
with the one of interpretation.\footnote{Although
a proponent of the epistemic view, Spekkens
suggests looking for ``the ontic states of
which quantum states are states of
knowledge''~\cite{spekkens}.  A specific way of
making sense of quantum mechanics
understood as a probability algorithm was recently
proposed by Mohrhoff~\cite{mohrhoff}.}

To address the third case, it will help focussing on
a particular argument along this line, the one
recently put forth by Bub~\cite{bub}.  Having
in mind mechanical extensions of the quantum theory
by means of nonclassical waves and particles, Bub
appeals to the following methodological principle:
\begin{quote}
[I]f $T'$ and $T''$ are empirically equivalent
extensions of a theory $T$, and if $T$ entails that,
in principle, there \emph{could not be} evidence
favoring one of the rival extensions $T'$ or $T''$,
then it is not rational to believe either $T'$ or
$T''$.
\end{quote}
Bub then contrasts the relationship between
the quantum theory and its mechanical extensions
like Bohmian mechanics, with the 
relationship between classical thermodynamics
and its mechanical explanation through the
kinetic-molecular theory.  In the latter case
there are empirical differences, as Einstein
showed in 1905 in the context of Brownian motion,
and as Perrin observed in 1909.

The realization that there may be empirical
differences between thermodynamics and the kinetic
theory of gases goes back to the
1860's~\cite{bader}, and is connected with the names
of Loschmidt and Maxwell.  But the kinetic theory is
much older.  Leaving aside the speculations of
the Greek atomists, we find atomism well alive
in the writings of Boyle and Newton in the seventeenth
and early eighteenth centuries.  D.~Bernouilli
explained the pressure of a gas in terms of atomic
collisions in 1738, and Dalton used
atoms to explain the law of multiple proportions
around 1808.  All this time though, there was little
indication that the empirical content of the atomic
models and the phenomenological theories might be
different.

One may argue whether it was rational
to believe in atoms before Loschmidt, but there
is no question that it was very fruitful.  As a
matter of fact, one of the reasons for
contemplating empirically equivalent extensions
of theories is that they may open the way
to nonequivalent theories.  This was clearly one
of Bohm's preoccupations in 1952~\cite{bohm1}.
That this perspective may or may not bear
fruits remains to be seen.

In the semantic view, the empirically equivalent
mechanical extensions $T'$ and $T''$ of $T$ are
rather called interpretations, emphasizing that
each points to how the world can possibly be the
way the theory says it is.  Equivalently, each shows
a way the theory can be true.  In classical
physics, finding ways that the theory can be true
is usually nonproblematic, as was illustrated
in Sec.~3.  This, however, is not the case with
the quantum theory.  Every logically consistent
interpretation proposed should therefore be viewed
as adding to the understanding of the theory.

Bub's methodological principle states that it is
not rational to believe in empirically equivalent
extensions $T'$ or $T''$ of $T$, if there cannot in
principle be evidence favoring the extensions.
Presumably, however, it is rational to believe in
$T$ (assuming empirical adequacy).  If $T$ is singled
out among its empirical equivalents, it must be
so on the basis of criteria other than empirical,
perhaps something like Ockham's razor.  This comes
as no surprise since even within the class of
internally consistent theories, acceptance almost never
depends on empirical criteria alone.

What criteria, other than empirical, make for
theory acceptance has been the subject of lively
debate, and the question may never be settled.
But the problem of acceptance translates to
interpretations.  Assume that a theory is
empirically adequate, and consider all the ways the
world can be for the theory to be so.  Each of
these ways is an interpretation.  Let an
interpretation be true just in case the world
is actually like it says it is.  Necessarily
then, an empirically adequate theory has one
of its interpretations that is true.

Now assume that, among all available
interpretations of a theory, no one is found
acceptable on a given nonempirical criterion.
Any proponent of the theory who believes
that the criterion is important is then faced
with the following choice.  Either he hopes that an
interpretation meeting the criterion
will eventually be found and, whether or not
he actually looks for it, he grants that the
search is an important task; or he concludes
that the theory, in spite of being empirically
adequate, is not acceptable.
%
%\newpage
\section{CONCLUSION}
In attempting to solve the problems of
measurement and instantaneous collapse,
the epistemic view asserts that state vectors
do not represent states of objects like
electrons and photons, but rather the
information we have on the potential
consequences of our experimental interventions
into nature.  I have argued that this picture
is adequate in a world where the theoretical
structure used in the prediction of these
consequences is independent of the one used
in the description of fundamental material
constituents.

In the world we live in, however, whatever
is responsible for clicks in Geiger counters
or cascades in photomultipliers also forms
the basis of material structure.  The
epistemic view can be construed as denying
that micro-objects exist or have states, or
as being agnostic about any logically
coherent connection between them and
macroscopic objects.  In the first case,
the solution it proposes to the foundational
problems is simply paradoxical, and calls for
an investigation of how it can be true.  In
the second case, it posits the existence of
a link between quantum and macroscopic objects.
Once this is realized, the urge to investigate
the nature of this connection will not easily
subside.
%
%\newpage

%
\end{document}